# How CO Affects the Composition of Titan's Tholins Generated with ECR Plasma

Zhengbo Yang[1], Yu Liu[1,2,3]*, Chao He[1]*, Pengcheng Yu[1], Rong Jin[1], Xiangqun Liu[1], Jinpu Zhang[1], Jiuhou Lei[1,2,3]

1.Deep Space Exploration Laboratory/School of Earth and Space Sciences, University of Science and Technology of China, No.96 Jinzhai Road, Hefei, 230026, Anhui, China

2.CAS Center for Excellence in Comparative Planetology/CAS Key Laboratory of Geospace Environment/Mengcheng National Geophysical Observatory, University of Science and Technology of China, No.96 Jinzhai Road, Hefei, 230026, Anhui, China

3.Collaborative Innovation Center of Astronautical Science and Technology, 92 Xidazhi Street, 150001, Heilongjiang, China

Corresponding Author: Yu Liu (yliu001@ustc.edu.cn), Chao He (chaohe23@ustc.edu.cn)

## Abstract

Titan's atmosphere possesses thick haze layers, but their formation mechanisms remain poorly understood, including the influence of oxygen-containing gas components on organic matter synthesis. As the most abundant oxygen-containing gas, the presence of CO has been found to exert a significant impact on the generation of oxygen-containing organic compounds. Therefore, investigating the influence of CO on the production and composition of Tholins through laboratory simulations, holds profound scientific significance in the context of Titan. The work presented here is an experimental simulation designed to evaluate the impact of CO on the atmospheric chemistry of Titan. To this end, CO was introduced into the standard $N_2/CH_4$ mixture at varying mixing ratios from 0.2% to 9%, and exposed to Electron Cyclotron Resonance (ECR) plasma to initiate photochemical reactions. Optical emission spectroscopy was employed for gas-phase in situ characterization, while infrared spectroscopy and high-resolution mass spectrometry were used to analyze the resulting solid products (tholins). Our results demonstrate that the addition of CO enriches the complexity of the chemical system. CO not only supplies oxygen to the system, but also enhances nitrogen's reactivity and incorporation, enhancing the number and quantity of the organic products.

# 1. Introduction

Titan is unique in our solar system: it is the only moon with a dense atmosphere, and the only solar system body besides Earth to boast a dense $N_2$ atmosphere. It harbors an extremely complex organic chemistry, and stable surface liquid. The atmosphere of Titan is primarily composed of $N_2$, with minor amounts of $CH_4$ and less than 1% $H_2$. Its surface temperature is approximately 94 K and, surface pressure is about 1.5 bar (Eshleman et al. 1983; Kunde et al. 1981). The pressure in the upper atmosphere is much lower, where photochemistry produces complex organic molecules and leads to the formation of haze particles (Fulchignoni et al. 2005; McKay et al. 1989). Titan's mildly reducing atmosphere is favorable for organic haze formation, and the presence of some oxygen-bearing molecules suggests that molecules of prebiotic interest may form in its atmosphere (Hörst et al. 2012; Sebree et al. 2018). The combination of liquid and organics means that Titan may be the ideal place in the solar system to test ideas about habitability, prebiotic chemistry, and the ubiquity and diversity of life in the universe (Cable et al. 2012).

The term 'tholins' was initially introduced by Sagan et al, to describe Titan atmospheric haze analogs obtained from methane-containing gas mixture in their laboratory discharge experiments (Sagan & Khare 1979). In the past decades, many research teams simulate various chemical processes in Titan's atmosphere by using different kinds of energy sources, such as UV lamps (Gupta et al. 1981), Laser Induced Plasma (LIP) (Borucki et al. 1988), high-energy spark (Scattergood et al. 1989), glow discharge (Coll et al. 1995), corona discharge (Navarro-González et al. 2001), high-frequency electrical discharge (Sarker et al. 2003), capacitance coupled plasma (CCP)(Perrin et al. 2021; Szopa et al. 2006), and radio frequency (RF) discharge (Reid Thompson et al. 1991). Not only the energy sources, but also many experimental setups with different innovative designs were employed to simulate the photochemistry of Titan's upper atmosphere, including Photochemical Flow Reactor (Clarke et al. 2000), the Laboratory Interuniversitaire des Systèmes Atmosphériques (LISA) (Coll, et al. 1995), the S.E.T.U.P. (French acronym for Experimental and Theoretical Simulation Useful for Planetology studies) program (Romanzin et al. 2008), the Titan Haze Simulation (THS) (Nuevo et al. 2022; Sciamma-O'Brien et al. 2019), the Production d'Aéroslos en Microgravité par Plasma REactifs (PAMPRE) (Perrin, et al. 2021; Szopa, et al. 2006), and the Planetary HAZE Research (PHAZER) (He et al. 2017; He et al. 2022). However, most of these devices primarily focus on the $N_2 - CH_4$ discharge reaction for hydrocarbon synthesis, with little consideration given to the role of oxygen-containing gas components in tholin synthesis, despite the fact that oxygen is essential for the synthesis of prebiotic molecules necessary for the origin of life.

Carbon monoxide, the most abundant oxygen-containing gas and the fourth most prevalent gas in Titan's atmosphere, is widely considered to significantly influence the formation of oxygenated organic compounds (de Kok et al. 2007; Yung et al. 1984). As early as 1982, Ishigami et al. conducted glow discharge experiments with CO-doped gas to assess the potential formation of amino acids in Titan's atmosphere (Ishigami 1982). Since the discovery of CO through the occultation of Titan in 1984 (Muhleman et al. 1984;

Yung, et al. 1984), several laboratory simulations have been performed to study the impact of $CO$ on the photochemistry of the upper atmosphere of Titan. For example, Bernard et al. (2003) investigated CO's effect by introducing gas mixture of $N_2/CH_4/CO$ (98%/1.99%/0.01%) into the LISA device and detected ethylene oxide as the main oxygenated compound (Bernard et al. 2003). Tran et al. analyzed both of the gas-phase and solid products of simulated haze after adding CO to the mixture of $N_2 - CH_4 - H_2 - C_2H_2 - C_2H_4 - HC_3N$. An absorption band corresponding to ketone groups was observed in the infrared spectrum, suggesting the formation of solid oxygen-containing organic compounds (Tran et al. 2008). In 2012, Hörst et al. investigated the discharge of $N_2/CH_4/CO$ mixture using the PAMPRE device, and identified amino acids and nucleoside bases in the solid products (Hörst, et al. 2012), and Fleury et al. addressed the efforts of CO on the gas phase reactivity (Fleury et al. 2014). In 2014, Hörst and Tolbert investigated how CO affects the size and number density of haze particles (Hörst & Tolbert 2014). The Planetary HAZE Research (PHAZER) device was employed to investigate the effects of CO on gas-phase products, particle density, and composition of Titan haze analogues (He, et al. 2017).

The previous research primarily emphasizes the generation of new products and phenomena resulting from the introduction of small proportions of gas. However, the impact of CO is still poorly understood, including the impact of CO on the $N_2 - CH_4$ chemical system, the mechanism of oxygen incorporation, the optical properties and chemical characteristic of O-containing haze particles. So, in this work, the atmospheric chemical process of Titan was simulated using Electron Cyclotron Resonance (ECR) discharge under different conditions in $N_2/CH_4/CO$ gas mixtures. During the experiments, in-situ optical emission spectroscopy was employed to monitor the reactive species inside of the reactor. Infrared spectroscopy and mass spectrometry were used to investigate the effect of CO on the composition of the solid-phase products.

## 2. Experimental Setup and Analysis Method

In this work, an ECR plasma was used to simulate the energy source in Titan's atmosphere. The ECR device utilizes the resonance interaction between electromagnetic waves and electrons undergoing cyclotron motion in a magnetic field, allowing electromagnetic waves with frequencies below the plasma frequency to propagate along the magnetic field lines into the plasma. This enables the generation of high-density plasma (~$10^{10}$ cm$^{-3}$) without the need for sophisticated microwave resonant cavities. The detailed parameter and properties of the ECR plasma can be found in our previous work (Liu et al. 2023; Zhang et al. 2019). The ECR plasma source consists of a microwave generator, transmission system, antenna, magnetic field coil, and vacuum chamber. The antenna is isolated from the plasma to avoid charge deposition in the plasma. At the same time, the current on the magnetic field coil can be adjusted to control the magnetic field induction intensity in the vacuum chamber.

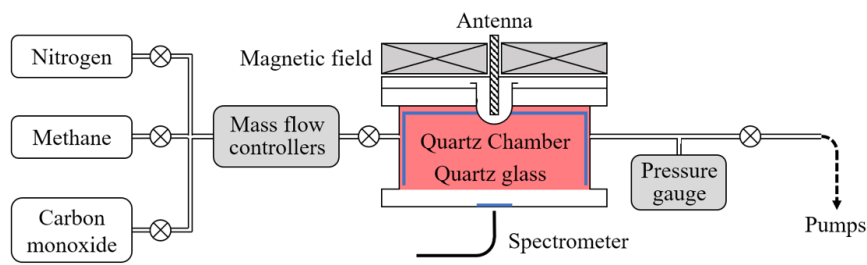

Figure 1. Schematic illustration of the experimental setup.

The experimental setup used in this work is illustrated in Figure 1, with the experimental conditions listed in Table 1. The gas mixture injected from the top of the vacuum chamber was controlled by digital mass flow controllers. The total gas flow rate is 50 standard cubic centimeters per minute (sccm), and the gas pressure is settled at 15 Pa. For all these experiments, $N_2$ (purity of 99.99%), $CH_4$ (purity of 99.99%) and CO (purity of 99.99%) were used as the working gases, and the gas mixture was discharged by the ECR plasma source (Input microwave power at 600W and magnetic field intensity at 875Gs). To explore the effect of the mixing ratio of $N_2/CH_4/CO$ on the formation of tholins, eleven groups of experiments were performed with varying mixing ratios. The first experiment only used $N_2$ as a reference. In the No.2-11 group of experiments, the ratio of CO was increased from 0% to 9% while the overall mass flow rate and the pressure in the chamber remained unchanged. In the No.1-3 group of experiments, we mainly focused on the effect of the addition of only different component gases on the experimental results.

Table 1. Experimental conditions at a fixed power (600 W)

| Group | Pressure/Pa | Mass flow/sccm | Ratio of gas ($N_2/CH_4/CO$) |
|---|---|---|---|
| 1 | 15 | 50 | 100%/0/0 |
| 2 | 15 | 50 | 95%/5%/0 |
| 3 | 15 | 50 | 94.8%/5%/0.2% |
| 4 | 15 | 50 | 94.6%/5%/0.4% |
| 5 | 15 | 50 | 94.4%/5%/0.6% |
| 6 | 15 | 50 | 94.2%/5%/0.8% |
| 7 | 15 | 50 | 94%/5%/1% |
| 8 | 15 | 50 | 92%/5%/3% |
| 9 | 15 | 50 | 90%/5%/5% |
| 10 | 15 | 50 | 88%/5%/7% |
| 11 | 15 | 50 | 86%/5%/9% |

The electron density of the ECR plasma changes with the composition of the gas mixture at a fixed power (Zhang et al. 2019). We measured the electron density of ECR plasma during the experiment using a Langmuir probe. As shown in Figure A1, the electron density changes with different gas mixtures. However, the density level remains within the same order of magnitude throughout the experiment, indicating the plasma density did not experience significant fluctuations (see Figure A1 in the Appendix). Optical emission spectroscopy (OES) was used for understanding the dissociation process of the mixture ($N_2/CH_4/CO$), which is one of the most important methods for studying the large variety of chemical processes occurring in plasma. A spectrometer (Horiba iHR550) was adopted to monitor the time-resolved OES. The plasma emission was directly guided to the spectrometer by an optical fiber collecting the light from in the plasma as the emission was sufficiently strong. The solid phase products (tholins), in the form of grains were collected and analyzed using ex-situ methods including infrared spectroscopy and mass spectrometry. For the infrared spectroscopy analysis, the solid samples were first mixed with KBr at the ratio of 1:20. After being ground and compressed by pellet presser at the press of 4 atmosphere (approximately 405 N/cm$^2$), the tholins/KBr pellets were prepared for transmission measurements using infrared spectroscopy (Nicolet 6700) from 500 to 4000 cm$^{-1}$. The spectra were acquired using deuterated triglycine sulfate (DTGS) detector with a resolution of 4 cm$^{-1}$. To resolve the extreme complexity of the samples, an ultrahigh resolution mass spectrometry was employed to determine the chemical composition of the Tholins. The samples were characterized using an LTQ-Orbitrap XL mass spectrometry, which has a high-resolution, accurate-mass (HRAM) in a mass range of 50-1000 Da.

## 3. Results and Discussion

*3.1 The in-situ OES results of the gas phase*

OES can be used to monitor the active substances in the reaction process, such as excited atoms, ions, and molecules, based on the analysis of emission bands and atomic lines. In this work, OES data was collected under different experimental conditions in the range of 300-800 nm (Figure 2). The typical luminescence spectra of $N_2$, $CH_4$, and CO at a pressure of 15 Pa and a fixed power of 600 W are shown in Figures 2a, 2b, and 2c, respectively. As shown in Figure 2a, various nitrogen emission peaks can be observed. The first positive system of excited $N_2$ ($B^3\Pi_g \rightarrow A^3\Sigma_u^+$) was identified in the range of 540-780 nm. Figure 2b shows the spectrum of $CH_4$ emissions, displaying two sharp and strong atomic spectral lines (the hydrogen Balmer lines: $H_\alpha$ at 656 nm, $H_\beta$ at 486 nm) with a few weak bands of molecular $H_2$ centered at 602 nm. The OES spectrum of CO is presented in Figure 2c, where it can be observed that the spectrum of CO presents relatively continuous spectral lines. This is because CO ionizes and produces free charged particles such as $CO^+$ and $O^+$, which can form various substances such as $CO_2$, $CO_2^+$, $O_2^+$, etc. under the collision of plasma. Due to the numerous reaction types involved, the spectrum of CO plasma is relatively continuous and has no obvious peaks.

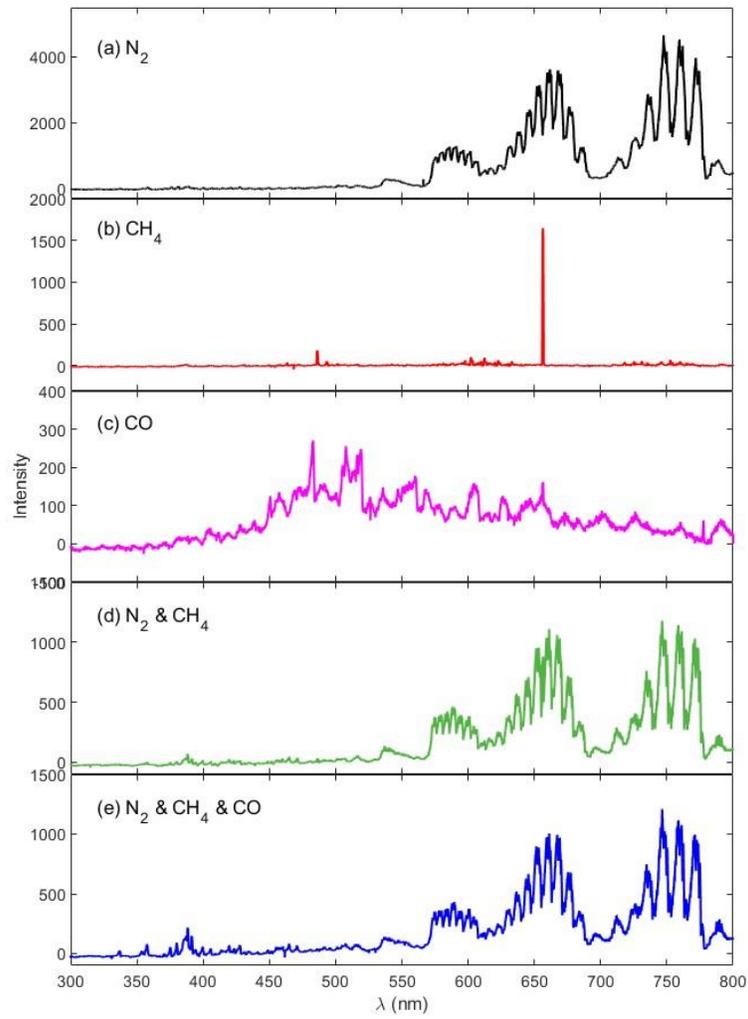

Figure 2. The OES spectra of (a) $N_2$; (b) $CH_4$; (c) CO; (d) mixed $N_2/CH_4$ ($CH_4$ at 5%) and (e) mixed $N_2/CH_4/CO$ (90%/5%/5%) at a pressure of 15 Pa at a fixed input power of 600 W.

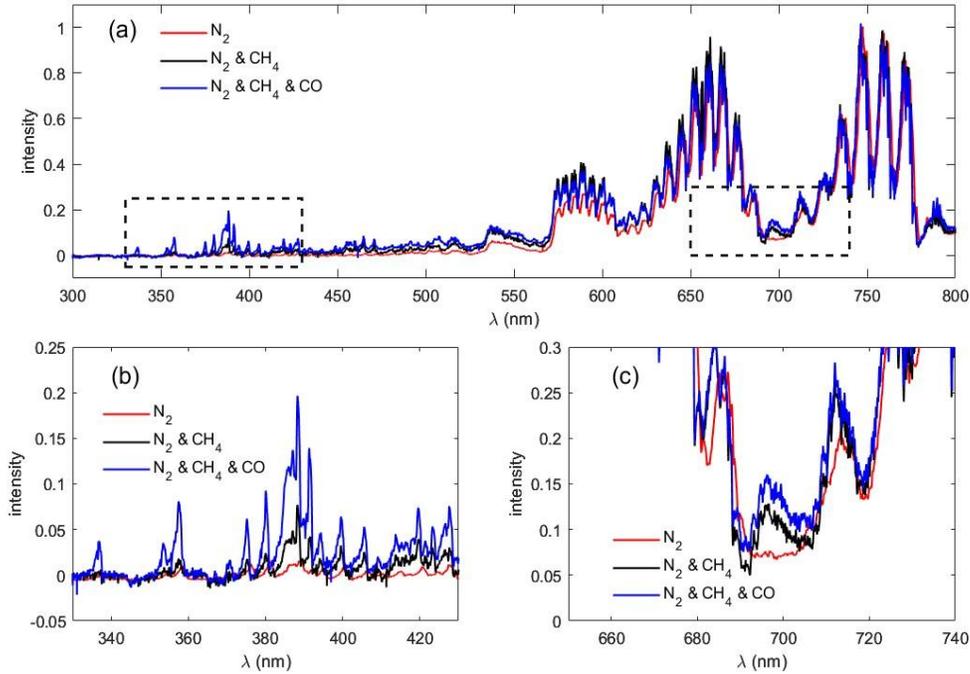

Figure 3. Comparison of the normalized OES spectra of $N_2$ (Figure 2a), mixed $N_2/CH_4$ (Figure 2d), and mixed $N_2/CH_4/CO$ (Figure 2e), where Figures 3b and 3c respectively represent the local OES spectra in the wavelength range of 330-430 nm and 650-740 nm.

The OES spectra of gas mixtures are shown in Figures 2d and 2e. These two spectra show the changes of peak and intensity with the addition of $CH_4$ and CO. The comparison of specific wavelengths is shown in Figure 3. Figure. 3a shows the normalized OES spectra of $N_2$, mixed $N_2/CH_4$, and mixed $N_2/CH_4/CO$. Peak enhancement is observed in several spectral regions (330-430 nm and 650-740 nm) with the addition of $CH_4$ (or $CH_4$ and CO). The specific wavelength ranges are shown in Figures 3b (330-430 nm) and 3c (650-740 nm). Signals that can be attributed to different nitrogen species, such as the narrow peaks centered at 337.1 and 380.0 nm, which correspond to the second positive system of excited $N_2$ ($C^3\Pi_u \rightarrow B^3\Pi_g$), and the $N_2^+$ peak (391.0 nm). As shown in Figure 3b, the relative intensity of the peaks (330-430 nm) increases after adding $CH_4$, and is further enhanced with the addition of CO, such as C≡N, identified at 388.3 nm. The peaks at 337 nm and 353/358 nm are only enhanced by the addition of CO not $CH_4$, indicating that CO promotes the formation of more reactive nitrogen species, which will be elaborated in detail in Section 3.3. Therefore, it can be considered that the addition of $CH_4$ and CO enhances the excitation level of the second positive band of $N_2$. For multi-component gas mixture, one component could affect the plasma density and gas excitation of the system through the mechanism as followed. The gas that requires lower energy for excitation (or ionization), yet possesses higher energy after being excited, which can further excite (or ionize) other harder-to-excite gas components. In our experiment, the excitation energy and bond dissociation energy of CO are lower than $N_2$, which facilitates its excitation (or ionization). The excited (or ionized) species from CO can provide sufficient energy and promote the excitation (or ionization) of nitrogen gas. What's more, the further enhancement of $N_2$ excitation in the second positive system probably increases nitrogen incorporation into larger molecules. The first positive system of excited $N_2$ ($B^3\Pi_g \to A^3\Sigma_u^+$) and second positive system of excited $N_2$ ($C^3\Pi_u \to B^3\Pi_g$) correspond to a series of spectral emissions resulting from electronic transitions between molecular orbitals. Among these, $C^3\Pi_u$, $B^3\Pi_g$, and $A^3\Sigma_u^+$ are significant metastable species (denoted as $N_2^*$). Compared to $A^3\Sigma_u^+$ (at 6.2 eV), the energy of excited stage of nitrogen is higher ($B^3\Pi_g$ at 7.3 eV and $C^3\Pi_u$ at 10.9 eV). The species at higher-energy-state, such as $C^3\Pi_u$, is capable of dissociating triple bond of $N_2$ (9.79 eV). The increase of the second positive system of $N_2$ indicates that more of higher-energy-state species ($C^3\Pi_u$) are produced in the system, therefore contributing more nitrogen reactivity.

Figure 3c clearly shows that the excitation of $N_2$ does not emit light in the range of 690-710 nm. However, after adding $CH_4$ and CO, new spectral lines appeared in this band. The addition of $CH_4$ to $N_2$ gas introduces C and H elements into the chemical system, which could lead to the formation of $N-H$ and $C\equiv N$. Therefore, the new band at 690-710 nm probably comes from $N-H$ emission, as the $C\equiv N$ is identified at 388.3 nm.

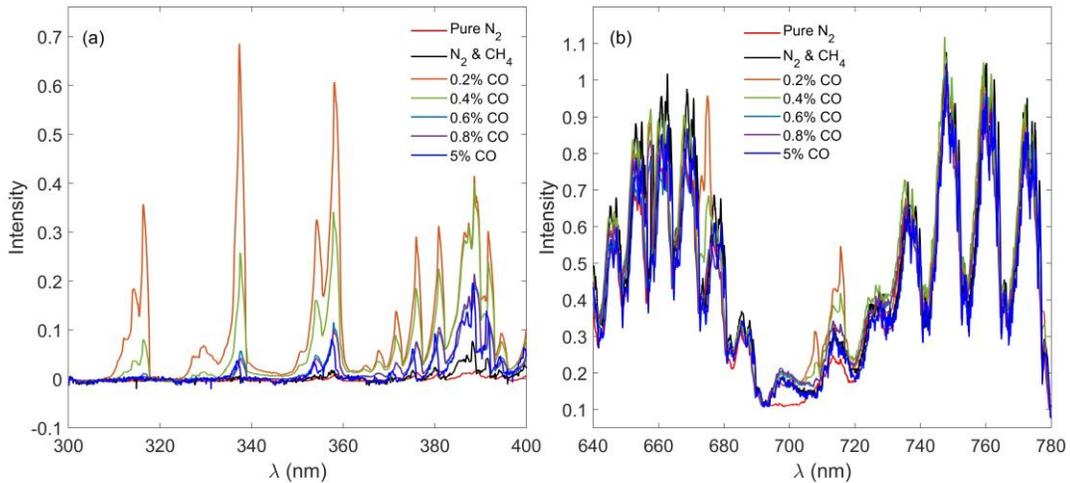

Figure 4. The OES spectra of pure $N_2$, $N_2/CH_4$ and different CO mixing ratios from 0.2% to 5% into mixed $N_2/CH_4$.

Figure 4 shows the normalized OES spectra of mixed $N_2/CH_4/CO$ with different CO ratios (0-5%). The spectral line intensity of this $N-H$ band (690-710 nm) did not change significantly with the increase of CO concentration (Figure 4b). However, figure 4a shows that the excitation level of $N_2$ first increases and then decreases with the rise in CO concentration, although it remains significantly stronger than that of the $N_2/CH_4$ case. The reason for less promotion of $N_2$ excitation (or ionization) with higher CO compared to lower CO is that, the continuous increase in CO leads to a higher frequency of both elastic and inelastic collisions between particles. Such collisions between ionized CO with other ions and electrons within the chamber cause plasma decay and subsequently lead to a reduction in ion density and electron temperature, which weakens the promotion of the excitation (or ionization) of $N_2$. In addition to the second positive system, the introduction of CO also effects the features at 675 nm, 708 nm, and 715.6 nm, which belong to Meinel band system and Herman infrared system derived from N, $N_2^+$, and $N_2^*$, respectively. The N, $N_2^+$, and $N_2^*$ can also contribute to the increase of nitrogen reactivity. The changing trend of these features are similar to that of other features of the second positive system from 300-400 nm, indicating that the addition of CO first increases and then decreases the degree of excitation/ionization.

### 3.2 The infrared spectroscopy analysis of the tholins

During the experiment, a yellow brown solid product deposits on the inner wall of the chamber. The sample was collected from a sampling plate (quartz glass) placed in the chamber. Subsequently, the tholins/KBr pellets were analyzed using infrared spectroscopy to investigate the absorption features of different functional groups in the solid. Figure 5 shows the infrared spectra of the tholins samples produced from $N_2/CH_4$ mixture and $N_2/CH_4/CO$ mixture. From the peak absorptions and intensities at different wavelengths, the characteristics of function groups in the tholins can be observed.

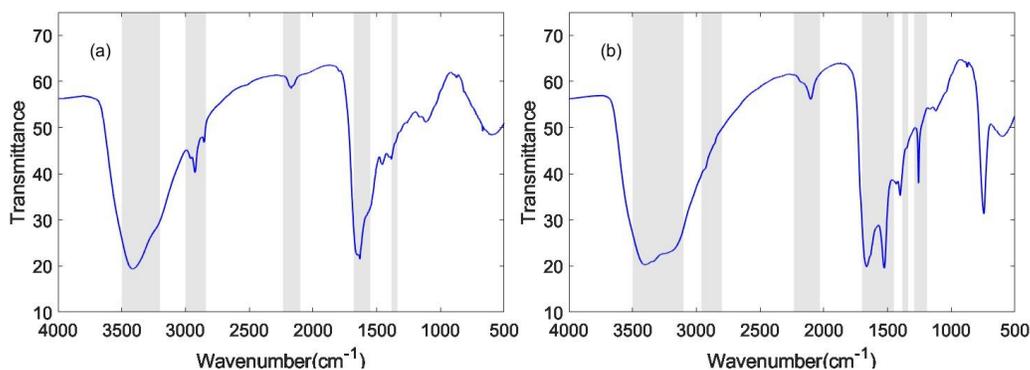

Figure 5. Comparison of the infrared spectra of mixed $N_2/CH_4$ at 95%/5% (a), and mixed $N_2/CH_4/CO$ at 90%/5%/5% (b).

For example, the $N-H$ band ($-NH_2$ or $-NH$, 3200-3500 cm$^{-1}$), $C-H$ stretching (2923 cm$^{-1}$ and 2960 cm$^{-1}$, respectively), $-CH_3$ bending mode (1338 cm$^{-1}$ or 1384 cm$^{-1}$), and $-C\equiv N/-N\equiv C/-N=C=N$ stretching mode (-2235 cm$^{-1}$) can be observed in Figure 5a. These results are generally consistent with previous studies on tholins (Bonnet et al. 2015; Dubois et al. 2019; Gautier et al. 2012; Imanaka et al. 2004). The wide absorption range of 3200 to 3500 cm$^{-1}$ indicates that hydrogen bonds are mainly dominated by N-H bonds, which is consistent with previous observations (Imanaka et al. 2012). However, the infrared spectra significantly change after adding CO, in which several new peaks appear compared to the one without CO (Bernard, et al. 2003). For example, the peak at 3300-3000 cm$^{-1}$ indicates more hydrogen bonded to unsaturated carbon in the CO sample, new peaks at 1526 cm$^{-1}$ comes from benzene ring skeleton, and the one at 1256 cm$^{-1}$ comes from CO vibrations in alcohols and/or ethers. In addition, the $O-H$ groups in carboxylic acid in the CO sample lead to a very broad peak from 2500-3500 cm$^{-1}$. Note that the cyano-group ($-$CN) stretching at 2110-2220 cm$^{-1}$ broads and splits into two peaks at 2160-2220 cm$^{-1}$ and 2060-2160 cm$^{-1}$, suggesting that more cyano-groups are bonded to unsaturated carbon and/or in aromatics with the addition of CO. Moreover, the peaks for saturated $CH_3/CH_2/CH$ at 2800-3000 cm$^{-1}$ decrease in the CO samples. Therefore, the addition of CO not only introduces oxygen atoms into the chemical system to form various oxygen-containing functional groups, but also facilitates the formation of unsaturated structures, such as benzene ring and aromatic structures, carbon-carbon double bonds, and carbon-carbon triple bonds. These observations will be corroborated in Section 3.3, which discusses the mass spectrometry results.

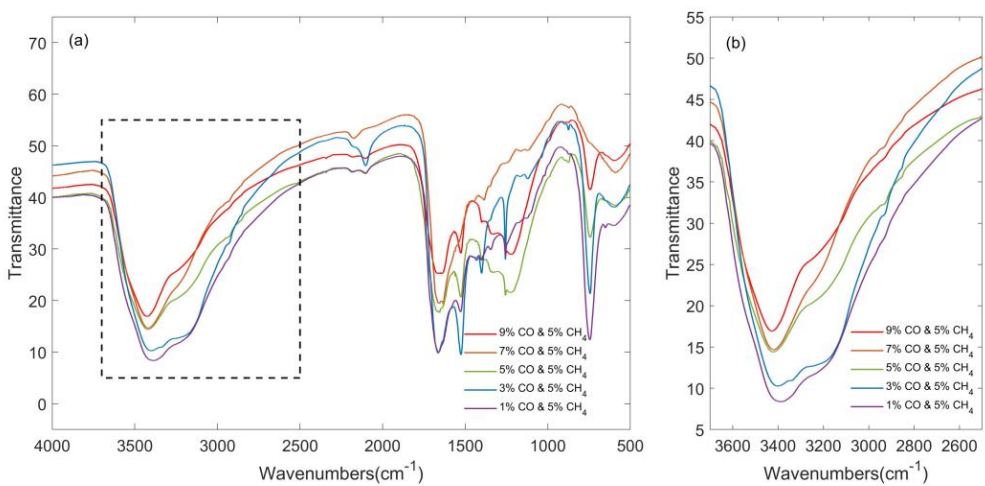

Figure 6. Infrared spectra of the $N_2/CH_4/CO$ mixture changing with the mixing ratio of CO in the gas mixture from 1% up to 9%

In addition, we also conducted infrared spectroscopy diagnosis of the tholins samples produced from the mixed gas with CO in different proportions (Figure 6a). The part of 3700-2500 cm$^{-1}$ is enlarged as shown in Figure 6b, and it can be observed that as the CO ratio gradually increases, the peak of this band shows a gradually blue-shift, shifting from 3385.4cm$^{-1}$ corresponding to 1% CO ratio to 3426.4 cm$^{-1}$ corresponding to 9% CO ratio. Both O $-$ H (3400-3300 cm$^{-1}$ in alcohol and 3500-2500 cm$^{-1}$ in carboxyl acid) and N $-$ H (3400-3250 cm$^{-1}$ in amines and 3500-3300 cm$^{-1}$ in amides) stretches appear in this wavelength region. Therefore, the observed blue-shift in the infrared spectrum is likely due to the formation of oxygen-containing functional groups, such as carboxylic acids and amides with increasing CO ratio in the initial gas mixture. Additionally, adding CO to the system changes the way how nitrogen atoms incorporate into molecular structure, possibly leading to the formation of amides. Such change could be related to the increased second positive system of nitrogen as shown in the OES spectra (Figure 3b). The peak blue-shift in this region has not been observed in prior experiments. Previous studies indicate that increasing the proportion of $CH_4$ without CO does not affect the infrared spectrum of this band (Dubois, et al. 2019), suggesting that the increase of $CH_4$ primarily contributes to the formation of nitrile compounds (Hörst et al. 2018). However, the introduction of CO into the system alters the mechanism by which nitrogen and hydrogen atoms combine to form N − H, ending up as amides.

### 3.3 The mass spectrometry results

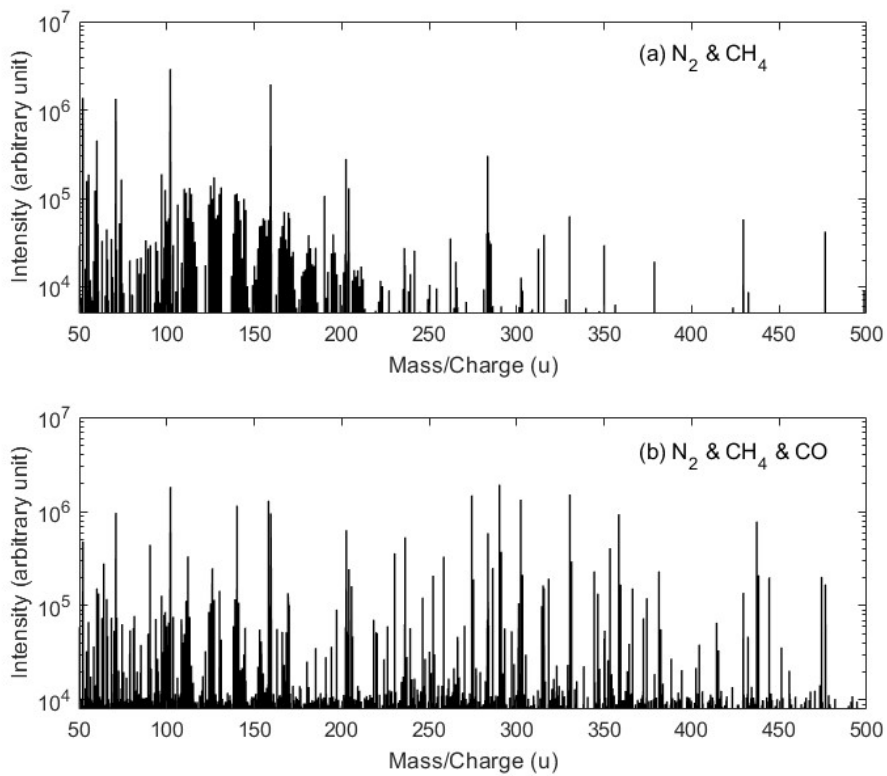

Figure 7. Orbitrap mass spectra in positive ion mode of the tholins produced from (a) $N_2/CH_4$ mixture (95:5) and (b) $N_2/CH_4/CO$ mixture (90:5:5). The intensity is shown in log scale.

In our study, ~85% of our samples are soluble in methanol. Previous studies already showed that Titan tholins are more soluble in polar solvent than in non-polar solvent (Carrasco et al. 2009; He & Smith 2014). Methanol ($CH_3OH$) was often used to dissolve the sample for MS measurements due to the relatively high solubility of the tholin samples in methanol. We added 5 ml solvent (methanol) into vial with 5 mg tholins in it. Given sufficient time for dissolution, the mixtures were filtered by 0.22 $\mu m$ syringe filter. After the solvent on the filter was evaporated, the mass of the insoluble solid was measured. The measurements of each sample were repeated for three times. The solubility of the tholins was calculated based on the remaining undissolved solid. The results revealed that the solubility of $N_2/CH_4$ sample is 83.38±6.7%, while the solubility of $N_2/CH_4/CO$ sample is 87.89±0.17%. The soluble portion was further analyzed with LTQ Orbitrap XL mass spectrometry in positive ion mode. Figure 7 shows that the mass spectra of tholins produced from $N_2/CH_4$ and $N_2/CH_4/CO$ gas mixtures. Both mass spectra are very crowded with thousands of peaks. Therefore, the soluble part of the solid product is a complex mixture, which is similar to the observations of other groups (Gautier et al. 2014; Moran et al. 2020). It is worth noting that with the $N_2/CH_4$ gas mixture, the mass peaks of resulting products show certain degree of periodicity. As shown in Figure 7a, obvious periodicity was observed between 50-250 amu, with a period of methyl (amu=15). After the addition of CO, new oxygen-containing groups and aromatic compounds are generated, resulting in the synthesis of complex macromolecules. Therefore, the mass spectrum (Figure 7b) shows relatively strong regularity in the small molecular weight range (amu<200), but the composition is more complex in the large molecular weight range. The formation of macromolecules indicates that the addition of CO enhances molecular diversity of the solid product and makes it possible to generate biomolecules.

The high-resolution mass spectra provide exact molecular mass, which allows us to calculate the molecular formula of each component in the complex products. By analyzing the data with a high-precision mass spectrometry, we identified several hundred formulas in the solid products. These compounds were classified into four subgroups based on their elemental compositions: CH, CHO, CHN, and CHON. We then performed statistical analyses on the abundance and variety of each subgroup, with results shown in Figure 8.

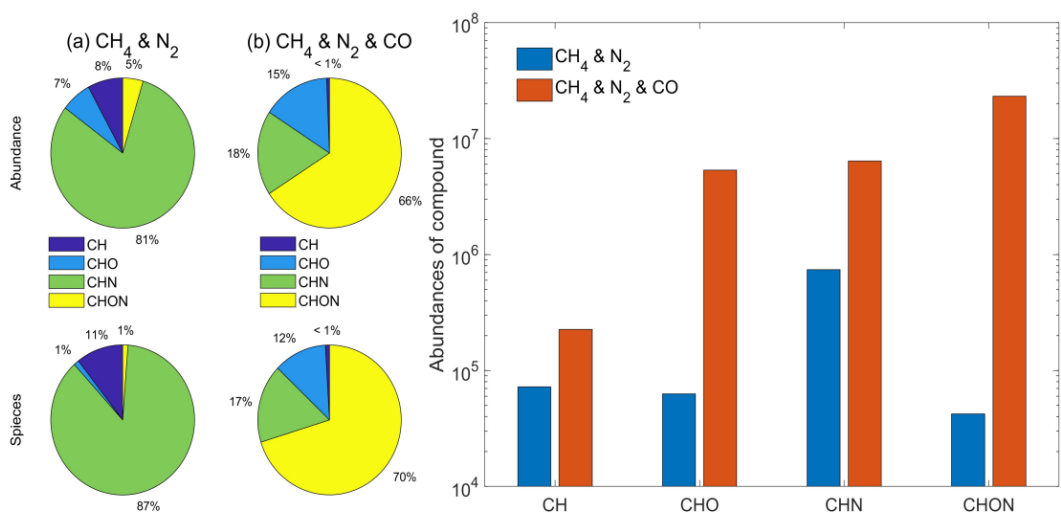

Figure 8. The pie charts on the left (Figure 8a) show the peak abundance percentage (the sum of the peak intensities of the matched within each subspecies) and species (the number of molecular types matched within each subspecies) percentage of each organic compound subgroup (i.e. CN, CHO, CHN) in the $N_2/CH_4$ sample, while the pie charts in the center (Figure 8b) show the peak abundance percentage and species percentage in the $N_2/CH_4/CO$ sample. The right figure (Figure 8c) shows a logarithmic comparison of peak abundances of organic compound subgroups for two samples.

From the Figure 8, it is evident that in the $N_2/CH_4$ sample, the predominant organic compounds are those with CH and CHN structures, with CHN compounds being the most prevalent. However, upon the addition of CO, the system synthesizes a greater quantity of organic compounds, particularly those containing oxygen, such as CHO and CHON, even the abundance of CHN-containing products was increased by over 100 times. This indicates that even a small amount of CO has a significant impact on the chemical system, notably accelerating the production of various organic compounds, not only oxygen-containing species, but also the nitrogen-containing ones.

The average chemical formulas were also calculated in both cases. In the $N_2/CH_4$ system, the average molecular formula was $C_{13.9953}H_{19.6006}O_{0.5501}N_{5.4265}$, whereas in the $N_2/CH_4/CO$ system, it was $C_{17.2066}H_{23.0388}O_{2.9677}N_{4.7956}$. Note that the small amount of oxygen in the $N_2/CH_4$ products could come from the oxygen and water contamination in the process of sample production and collection. However, the average molecular formula clearly shows a significant increase in average molecular weight (272.484 amu in the $N_2/CH_4$ system and 344.320 amu in the $N_2/CH_4/CO$ system), suggesting the production of larger organic molecules. Additionally, the overall oxygen content increased substantially, due to the formation of more oxygen-containing organic compounds such as carboxylic acids and amides. Furthermore, the system's degree of unsaturation also increased, indicating the formation of more double and triple bonds.

For the identified molecular formulas, we analyzed the relationships between the number of carbon atoms, the degree of unsaturation, and the number of nitrogen or oxygen atoms. The results are illustrated in Figure 9.

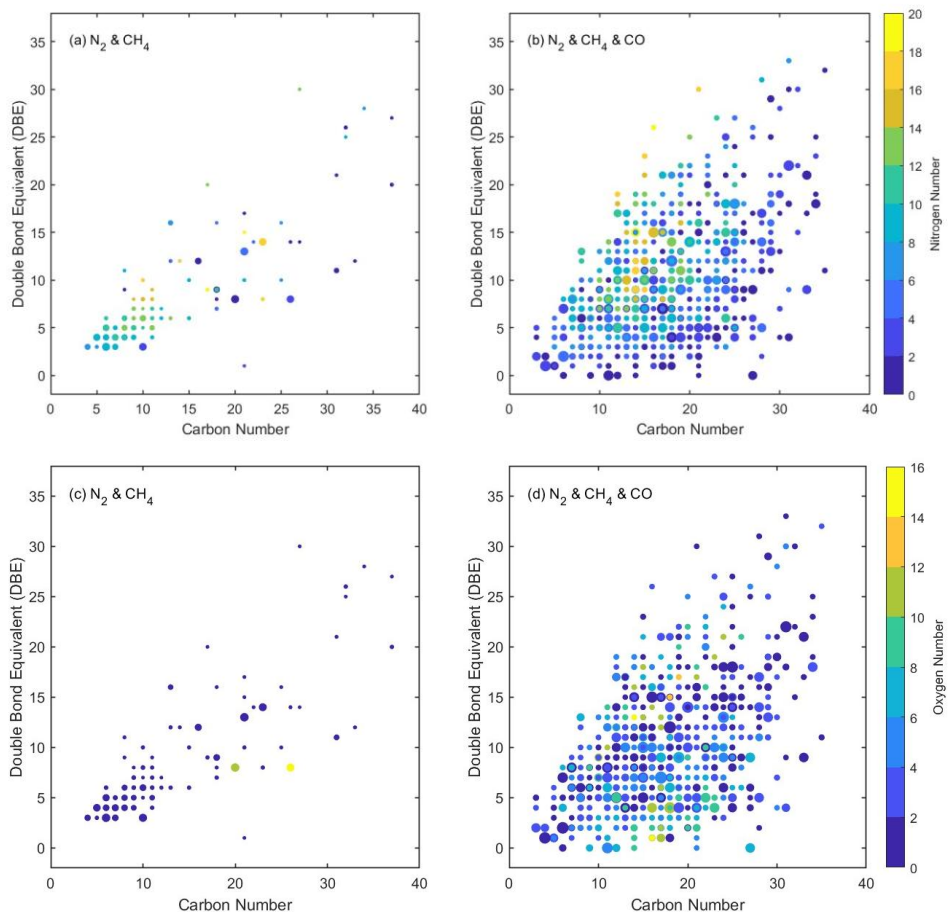

Figure 9. Carbon number vs. double bond equivalent (DBE) for organic species in the $N_2/CH_4$ sample (Figures 9a and 9c) and $N_2/CH_4/CO$ sample (Figures 9b and 9d). The color bar denotes the number of N atoms (Figures 9a and 9b) and O atoms (Figures 9c and 9d). The size of the symbols reflects the relative peak intensities of molecular formulae on a logarithmic scale. The Double Bond Equivalent (DBE) = $\#C - \frac{\#H - \#N}{2} + 1$, where $\#C$, $\#H$, and $\#N$ are number of carbon, hydrogen, and nitrogen atoms, respectively.

On a broader scale, the addition of small amount of CO significantly increased the complexity of the chemical system, leading to the synthesis of a greater variety and quantity of organic compounds. Comparing Figures 9a and 9b, it is evident that organic compounds with higher nitrogen atom counts are predominantly found within a carbon atom range of 10 to 20, with a degree of unsaturation greater than 7. A higher number of carbon atoms indicates the synthesis of larger organic molecules, and the higher degree of unsaturation is partly attributed to the presence of nitrile and amide products. This observation corroborates earlier inferences from OES and IR spectra, suggesting that the introduction of CO not only facilitates nitrogen excitation but also accelerates the incorporation of nitrogen atoms into the chemical system, increasing the production rate of nitrogen-containing organic compounds.

In comparing Figures 9c and 9d, it is observed that organic compounds with higher oxygen atom counts are mainly concentrated within a carbon atom range of 10 to 20, with degrees of unsaturation either very low

(approximately 0) or very high (approximately equal to the number of carbon atoms). This pattern suggests that oxygen is incorporated into the structures as either hydroxyl or carboxyl groups. Figures 9b and 9d reveal the presence of amides, carboxylic acids, nitrile, and more unsaturated structures in the $N_2/CH_4/CO$ sample, consistent with the IR results (Section 3.2). The addition of CO increases the degree of unsaturation in the solid product, which is also observed previous studies (He, et al. 2017; Moran et al. 2022). However, our study here is the first to demonstrate that introducing small amount of CO into the system can increase the complexity and diversity of the solid products. Note that the compositional information of the solid products is drawn from the comparison of the samples produced with 5% CO and without CO. This could hold true with lower CO mixing ratio because the gas phase OES spectra show a clear changing trend with the increase of CO. We will further verify whether lower CO mixing ratio have the same impact on the composition of the solid products in our follow-up study.

Table 2: Molecular formulas matched to proteinogenic and non-proteinogenic amino acids in tholins

| Number | Chemical formula | Name | Abundant |
|---|---|---|---|
| 1 | $C_3H_6O_2N_2$ | cycloserine | $1.28 \times 10^5$ |
| 2 | $C_4H_{10}O_2N_2$ | 2,4-Diaminobutanoic acid | $1.00 \times 10^6$ |
| 3 | $C_6H_{14}O_2N_4$ | Arginine | $1.31 \times 10^6$ |
| 4 | $C_7H_{13}O_2N_1$ | cyclohexanecarboxylic acid,1-amino | $2.18 \times 10^4$ |
| 5 | $C_8H_{15}O_2N_1$ | Cyclohexylglycine | $1.34 \times 10^4$ |
| 6 | $C_{10}H_{11}O_2N_3$ | Tryptazan | $1.01 \times 10^4$ |
| 7 | $C_{11}H_{12}O_2N_2$ | Tryptophan | $2.44 \times 10^5$ |
| 8 | $C_{26}H_{43}O_6N_1$ | glycocholic acid | $1.05 \times 10^4$ |

Additionally, previous experiments have revealed that prebiotic molecules (amino acids and nucleobases) can be formed in Titan's atmosphere through photochemical processes, and CO plays an important role in producing these species (Hörst, et al. 2012; Sebree, et al. 2018). Therefore, we searched the molecular formula of potential prebiotic species using the list described in Moran et al. (2020) and found several molecular formulas matching with proteinogenic and non-proteinogenic acids as listed in Table 2. The abundances of these species are relatively high, all above $1 \times 10^4$ as revealed in the mass spectrum. Note that the detection of the prebiotic molecular formulas does not guarantee that these species are formed in our experiment. Further follow-up work is required to determine whether these prebiotic species are indeed present in our sample.

## 4. Conclusion

In this study, the ECR discharge plasma was employed to investigate the production of tholins mimicking Titan's atmospheric haze. We monitored the gas phase composition during the experiment with OES and analyzed the solid product with IR spectroscopy and high-resolution mass spectrometry. OES shows that adding $CH_4$ to $N_2$ gas results in the formation of CN and NH, and promotes the excitation of $N_2$ in the second positive system; introducing CO to the $N_2/CH_4$ gas further enhances the excitation of $N_2$ in the second positive system and increases the production of CN but not NH. Infrared spectroscopy indicates that with the increase in CO ratio, a significant number of oxygen-containing functional groups, such as hydroxyl and carboxyl groups, appeared, and aromatic compounds as well as other more unsaturated structures were also detected. Mass spectrometry data reveal that the addition of CO greatly enhances the synthesis of macromolecular products, as evidenced by a dramatic increase in both the abundance and variety of organic compounds. The data from the mass spectrometer confirm the increase in cyanide content and the presence of amide and oxygen-containing groups (hydroxyl or carboxyl groups). We matched several prebiotic molecular formulas with higher relative abundance in the mixture, suggesting that atmospheric photochemical reactions could be a source for generating biochemical precursors. Our results demonstrate that adding small amount of CO into $N_2/CH_4$ changes the chemical system significantly, enhancing the reactivity of the gas species and increasing the complexity and diversity of the solid products. CO not only directly supplies oxygen to the system and leads to formation of oxygen-containing species as hydroxyl or carboxyl groups, but also delivers less-reduced carbon species that increase the degree of unsaturation of the solid products. Our study here reveals the important role of CO in chemical processes in both gas and solid phase, indicating that CO can dramatically impact the chemistry happening in Titan, Triton, and Pluto as well as other CO -containing planetary atmospheres. Future work is required to understand the role of CO in the formation of prebiotic species.

## Acknowledgments

This work was supported by the National Natural Science Foundation of China (42322407, 42188101, 92271105, 12305229), the National Key R&D Program of China (2022YFF0503702), the Project of Stable Support for Youth Team in Basic Research Field, CAS (YSBR-018), and the Youth Innovation Promotion Association CAS (2020451)

## Appendix

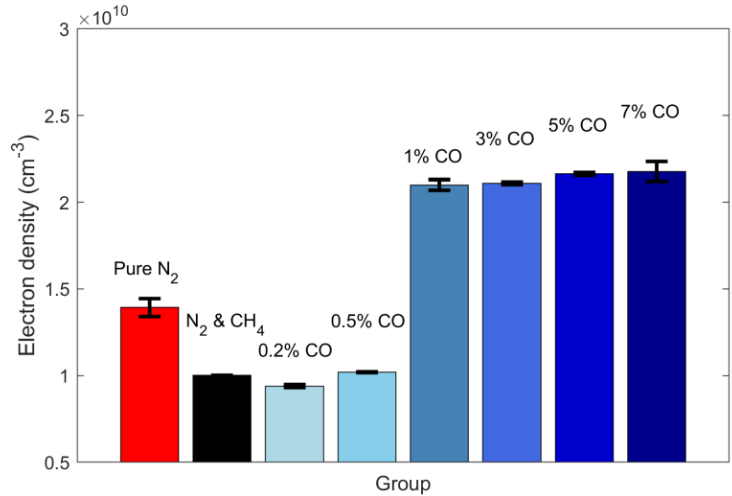

Figure A1. The electron density in different gas mixture measured with a Langmuir probe. The gas mixtures are 5% $CH_4$ with CO mixing ratio increasing from 0% to 7%. The error bar is derived from multiple measurements for each case.

Based on the measured electron density and experimental condition, we calculate the ECR-induced ionization degree ($\alpha$) of the gas mixture using the following equation

$$\alpha = \frac{n_e V}{n_{total}} = \frac{n_e V}{\frac{pV}{RT}} = \frac{n_e RT}{p},$$

where $n_e V$ and $n_{total}$ are the amount of electron and gas molecules in the chamber. $n_e$ is the electron density (cm$^{-3}$). The calculated ionization degree in the 5% CO experiment is about 5.8×10$^{-6}$, indicating that the ratio of ions and electrons in the chamber is extremely low during the discharge process, and the majority of the components remain as neutral gas species.